\PassOptionsToPackage{dvipsnames,table}{xcolor}
\documentclass[sigconf]{acmart}
\renewcommand\footnotetextcopyrightpermission[1]{} 
\usepackage{multirow}
\usepackage{booktabs}
\usepackage{graphicx}
\usepackage{array}
\usepackage[most]{tcolorbox}
\usepackage{colortbl}
\usepackage{amsmath}
\usepackage{xspace}
\usepackage{enumitem}
\usepackage{hyperref}
\usepackage{listings}
\usepackage{pifont}
\usepackage{soul}
\usepackage{url}
\usepackage{subfigure}
\usepackage[ruled,vlined]{algorithm2e}
\usepackage{tabularx}
\usepackage{caption}
\usepackage{fontawesome5}
\usepackage{balance}

\definecolor{LemonChiffon}{RGB}{255, 250, 205}
\definecolor{LightSteelBlue}{RGB}{176, 196, 222}
\definecolor{LightCoral}{RGB}{240, 128, 128}
\definecolor{LightCyan}{RGB}{224, 255, 255}

\newtcolorbox{finding}{
  colback=LemonChiffon,
  colframe=orange!60!black,
  boxrule=0pt,
  sharp corners,
  boxsep=1mm,
  left=1mm,   
  right=1mm,  
  top=1.5mm,    
  bottom=1.5mm  
}

\newtcolorbox{promptbox}{
  colback=black!5,          
  colframe=black!60,        
  boxrule=0.5pt,
  sharp corners,
  boxsep=1.5mm,
  left=2mm,
  right=2mm,
  top=2mm,
  bottom=2mm,
}

\tcbset{
  promptbox/.style={
    enhanced,
    colback=gray!5,
    colframe=gray!50,
    boxrule=0.5pt,
    arc=2mm,
    outer arc=2mm,
    fonttitle=\bfseries,
    coltitle=black,
    title={#1},
    sharp corners=south,
    breakable,
    listing only,
    listing options={basicstyle=\ttfamily\footnotesize, breaklines=true}
  }
}

\newcommand{\approach}{{SWE-Debate}\xspace} 

\newcommand{\lparagraph}[1]{\textbf{#1}~}

\newcommand\blfootnote[1]{%
  \begingroup
  \renewcommand\thefootnote{}\footnote{#1}%
  \addtocounter{footnote}{-1}%
  \endgroup
}

\newcolumntype{L}[1]{>{\raggedright\arraybackslash}p{#1}}
\newcolumntype{Y}{>{\raggedright\arraybackslash}X}

\newtcbox{\hlprimarytab}{on line, rounded corners, box align=base, colback=c3!10,colframe=white,size=fbox,arc=3pt, before upper=\strut, top=-2pt, bottom=-4pt, left=-2pt, right=-2pt, boxrule=0pt}
\newtcbox{\hlsecondarytab}{on line, box align=base, colback=blue!10,colframe=white,size=fbox,arc=3pt, before upper=\strut, top=-2pt, bottom=-4pt, left=-2pt, right=-2pt, boxrule=0pt}

\definecolor{c3}{cmyk}{0.3081,0,0.7209,0.3255} 

\author{Han Li\textsuperscript{\textdagger}}
\affiliation{%
 \institution{Shanghai Jiao Tong University}
 \country{China}}
\email{lihan0421@sjtu.edu.cn}

\author{Yuling Shi\textsuperscript{\textdagger}}
\affiliation{%
 \institution{Shanghai Jiao Tong University}
 \country{China}}
\email{yuling.shi@sjtu.edu.cn}

\author{Shaoxin Lin}
\affiliation{%
 \institution{Huawei}
 \country{China}}
\email{2120200411@mail.nankai.edu.cn}

\author{Xiaodong Gu\textsuperscript{$\ddag$}}
\affiliation{%
 \institution{Shanghai Jiao Tong University}
 \country{China}}
\email{xiaodong.gu@sjtu.edu.cn}

\author{Heng Lian}
\affiliation{%
  \institution{Xidian University}
  \country{China}}
\email{lianheng23@163.com}

\author{Xin Wang}
\affiliation{%
  \institution{Huawei}
  \country{China}}
\email{betterwangx@foxmail.com}

\author{Yantao Jia}
\affiliation{%
  \institution{Huawei}
  \country{China}}
\email{jamaths.h@163.com}

\author{Tao Huang}
\affiliation{%
  \institution{Huawei}
  \country{China}}
\email{276255565@qq.com}

\author{Qianxiang Wang}
\affiliation{%
  \institution{Huawei}
  \country{China}}
\email{wangqianxiang@huawei.com}

\title{\approach: Competitive Multi-Agent Debate for Software Issue Resolution}

\begin{document}


\begin{abstract}
    \blfootnote{\textdagger Equal contribution.}\blfootnote{\textsuperscript{$\ddag$}Xiaodong Gu is the corresponding author.}Issue resolution has made remarkable progress thanks to the advanced reasoning capabilities of large language models (LLMs). Recently, agent-based frameworks such as SWE-agent have further advanced this progress by enabling autonomous, tool-using agents to tackle complex software engineering tasks. 
    While existing agent-based issue resolution approaches are primarily based on agents' independent explorations, they often get stuck in local solutions and fail to identify issue patterns that span across different parts of the codebase. 
    To address this limitation, we propose \approach, a competitive multi-agent debate framework that encourages diverse reasoning paths and achieves more consolidated issue localization.
    \approach first creates multiple fault propagation traces as localization proposals by traversing a code dependency graph. Then, it organizes a three-round debate among specialized agents, each embodying distinct reasoning perspectives along the fault propagation trace. This structured competition enables agents to collaboratively converge on a consolidated fix plan. 
    Finally, this consolidated fix plan is integrated into an MCTS-based code modification agent for patch generation. Experiments on the SWE-bench benchmark show that SWE-Debate achieves new state-of-the-art results in open-source agent frameworks and outperforms baselines by a large margin\footnote{Our code and data are available at \url{https://github.com/YerbaPage/SWE-Debate}}. 
\end{abstract}


\maketitle

\section{Introduction}

Automated repository-level issue resolution has emerged as a critical challenge in software engineering. The task aims to automatically localize and fix the defective code snippets, based on reported issues. 
In software development, developers spend a majority of their debugging efforts in understanding code and making changes~\cite{jimenez2024swebench,zhang2024autocoderover}. 
Meanwhile, automated tools often struggle with the same challenge~\cite{chen2024whena,wang2025are,shao2025llms}. 
Inadequate code understanding leads to incomplete fixes, introduces new bugs, and significantly extends development cycles~\cite{chen2025unveilinga,cuadron2025danger}. 

The key challenge in effective issue resolution is fault localization, namely, identifying the code snippets triggering the specific issue~\cite{xia2024agentless}.
Unlike conventional code retrieval, fault localization requires a deeper connection between natural language issue descriptions and programming language structures. This process requires reasoning over the structural and semantic properties of code, often across complex dependency graphs~\cite{liu2024marscode,chen2025locagent,jiang2025cosil}, and entails a comprehensive understanding of software architecture as well as strategic decision-making.

The emergence of LLMs has significantly advanced this area by leveraging code understanding and reasoning capabilities~\cite{yang2024large,wu2023large,shi2024between}. More recently, agent-based methods~\cite{yang2024sweagenta,zhang2024autocoderover,antoniades2024swesearch,wang2024openhands,chen2024coder} have emerged, simulating autonomous agents capable of tool use and high-level decision-making. 
These approaches use iterative exploration and planning to enable systematic codebase traversal, representing a shift toward structured and interactive issue resolution processes~\cite{ma2025improving,xia2024agentless}.

While agent-based approaches have shown notable progress on standard benchmarks such as SWE-bench~\cite{jimenez2024swebench}, they mainly rely on agents' independent exploration, that is, the agents individually understand code repository and propose their modification plans. As a result, they often get stuck in local solutions and fail to identify issue patterns that span across large, complex codebases~\cite{chen2025unveilinga,cuadron2025danger}. This fundamental limitation stems from a core challenge we term \textit{limited observation scope}
~\cite{qin2025agentfl,yu2025orcalocab}: when multiple code locations appear relevant to the issue description, correct resolution often depends on a deep understanding of code structure and component relationships. However, independent exploring agents lack the diverse analytical perspectives needed to systematically compare and rank these competing alternatives. This limitation becomes more pronounced when agents are faced with multiple plausible fix strategies or modification points, each with different implications for maintainability, compatibility, and architectural soundness~\cite{chen2025locagent,jiang2025cosil,zhang2024autocoderover,liu2024marscode}. Without sufficient reasoning capacity to evaluate these trade-offs holistically, independent exploring agents often fail to directly identify the correct fix location and strategy, relying instead on repeated trial-and-error that is both inefficient and error-prone~\cite{antoniades2024swesearch,ehrlich2025codemonkeys}.

\begin{figure*}[t]
    \centering
    \includegraphics[width=0.9\textwidth]{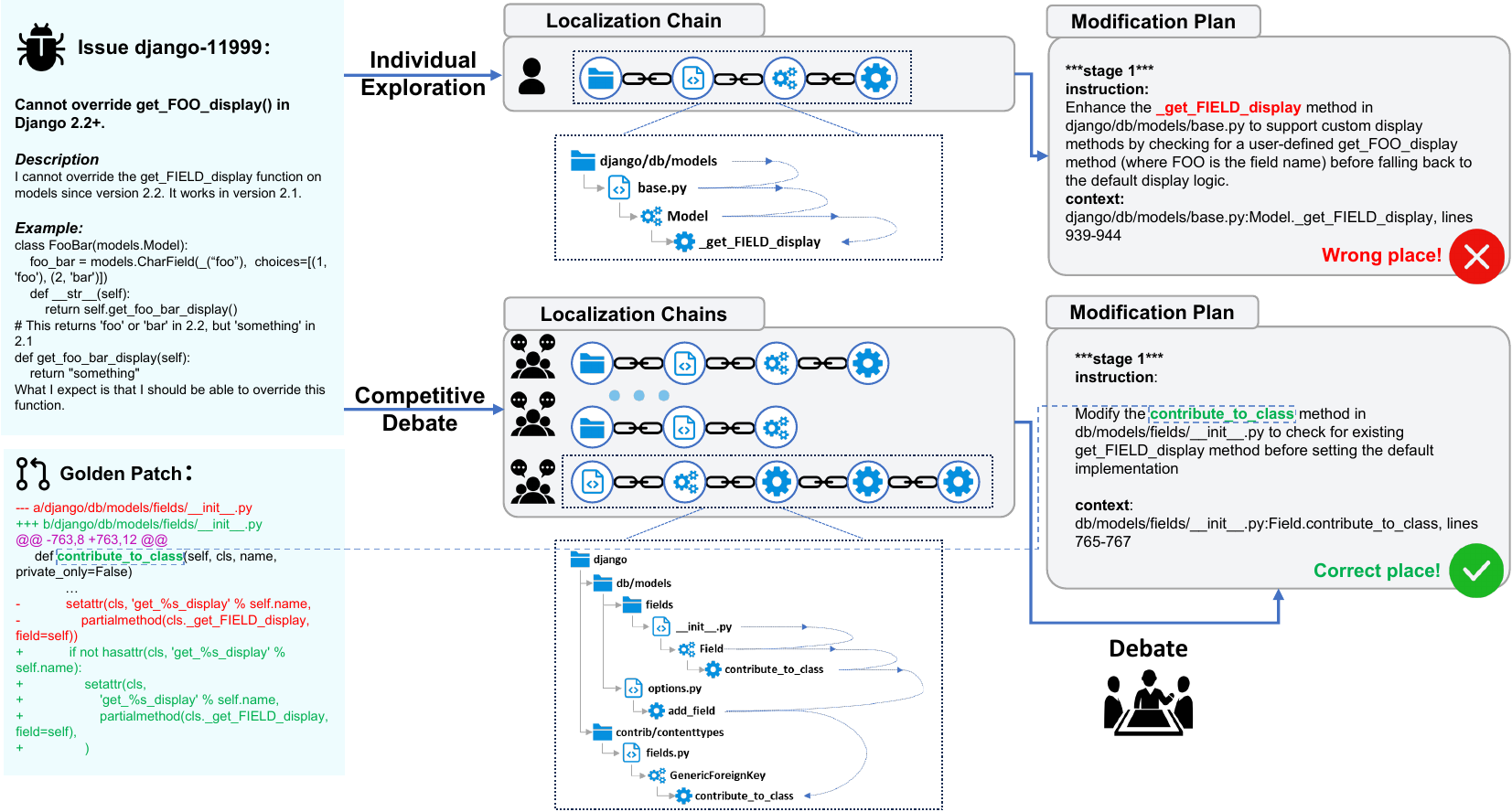}
    \caption{Motivating example of multi-agent debate.}
    \label{fig:motivation}
\end{figure*}

To address these challenges, we propose \approach, a competitive multi-agent debate framework that promotes diverse reasoning paths and achieves more consolidated fault localization. \approach reframes issue resolution through graph-guided localization and structured debate mechanisms. 
The framework operates through a three-stage pipeline. 
First, it creates multiple fault propagation traces as localization proposals by dependency analysis across the codebase. Specifically, a static dependency graph is built to represent relationships among code entities—such as function calls, class inheritance, module imports, and variable references. Using language model-based semantic matching, \approach identifies entities that are most relevant to the issue description, which serve as high-confidence entry points for chain construction. \approach traverses the graph from each entry point, 
yielding a set of candidate localization chains. Each chain captures a potential fault propagation path, reflecting different structural viewpoints, i.e., alternative code organization contexts in which the issue may appear, such as along a call hierarchy, inheritance structure, or shared data flow. 
Next, the algorithm creates a consensus fix plan through a structured three-round debate process. In the first round, multiple agents engage in competitive ranking to select the most promising fault propagation trace. Based on the selected trace, agents independently propose candidate modification plans based on different reasoning perspectives, then engage in competitive refinement to defend their proposals while critiquing alternatives. A discriminator selects the most promising plan, synthesizing insights from the debate to produce a coherent and actionable modification plan. 
In the final stage, the modification plan is used to initialize a Monte Carlo Tree Search (MCTS) framework~\cite{antoniades2024swesearch} for patch generation. 

Our experimental evaluation on the SWE-Bench-Verified dataset systematically compares \approach against state-of-the-art baselines. \approach achieves new state-of-the-art results under open-source agent frameworks and outperforms baseline methods by a large margin. 
Ablation studies show that the multiple chain generation mechanism provides the largest contribution to overall performance, validating our hypothesis that the fault propagation traces proposal enables more accurate fault localization and issue resolution. 

Our main contributions include: 
\begin{itemize}[leftmargin=0.5cm]
    \item A novel method to generate multiple candidate fault propagation traces. The method captures diverse potential fault propagation paths through code dependencies and structural relationships.
    \item A competitive multi-agent debate framework for precise fault localization through diverse reasoning perspectives and structured argumentation.
    \item Extensive experiments show that our competitive debate paradigms achieve 6.7\% improvement in issue resolution rate and 5.1\% improvement in fault localization accuracy.
\end{itemize}

\section{Motivation}

Repository-level issue resolution reveals the fundamental limited observation scope that agentic approaches face in complex software engineering scenarios~\cite{chen2025unveilinga,cuadron2025danger}. While individual agents can successfully handle straightforward localization tasks, they systematically fail when multiple code locations appear relevant to issue descriptions, requiring comprehensive architectural understanding and careful evaluation of competing modification plans~\cite{qin2025agentfl,yu2025orcalocab}. The core challenge lies in the inherent perspective limitations that prevent single agents from accurately assessing the trade-offs between multiple viable solutions~\cite{du2024improving,chan2023chateval}.

\lparagraph{The Individual Exploration Problem} As illustrated in Figure~\ref{fig:motivation}, single-agent individual exploitation approaches rely on individual exploration where agents independently understand code repositories and propose modification plans without systematic evaluation of alternative approaches. Consider the Django-11999 issue where users cannot override \texttt{get\_\allowbreak FOO\_\allowbreak display()} methods in Django 2.2+. The isolated agent performs semantic search with query \texttt{"get\_\allowbreak FOO\_\allowbreak display impl"} and immediately focuses on \texttt{django/\allowbreak db/\allowbreak models/\allowbreak base.\allowbreak py}, specifically the \texttt{\_get\_\allowbreak FIELD\_\allowbreak display} method. This individual exploration path appears reasonable from a single perspective but represents a fundamental misunderstanding of the issue's root cause.

The individual exploration approach exemplifies the core limited observation scope problem~\cite{chen2025locagent,jiang2025cosil}: the agent cannot systematically evaluate whether the base method implementation or the field registration mechanism contains the actual fault source~\cite{qin2025agentfl}, lacks the diverse reasoning perspectives necessary to compare runtime workarounds against structural solutions~\cite{zhang2024autocoderover,liu2024marscode}, and cannot effectively analyze the architectural trade-offs between different modification plans~\cite{antoniades2024swesearch,ehrlich2025codemonkeys}. This individual exploration limitation prevents the agent from reconsidering its fundamental localization strategy when initial approaches fail, leading to inefficient trial-and-error cycles that characterize single-agent individual exploration methods.

\lparagraph{Multi-Agent Debate Resolution} The correct resolution requires recognizing that method override failures stem from the field registration process in \texttt{Field.\allowbreak contribute\_\allowbreak to\_\allowbreak class}, where Django unconditionally overwrites user-defined methods during class construction. This structural insight emerges through multi-agent debate where different agents examine alternative code regions and systematically defend their localization strategies against competing interpretations, moving beyond individual exploration to structured competitive analysis~\cite{du2024improving,chan2023chateval}.

In the Django-11999 case, multi-agent debate directly addresses the limited observation scope: Agent A advocates for modifying the base \texttt{\_get\_\allowbreak FIELD\_\allowbreak display} method while Agent B argues for intervention in \texttt{Field.\allowbreak contribute\_\allowbreak to\_\allowbreak class}, forcing systematic comparison of these structurally different approaches. Through structured debate, different agents defend competing modification philosophies—runtime flexibility versus source-level prevention—revealing architectural trade-offs and maintainability implications that are invisible to individual exploration. When initial approaches fail, competitive pressure prevents agents from abandoning promising directions and instead drives systematic analysis of why specific strategies succeed or fail, transforming individual exploration into collaborative reasoning that enables comprehensive evaluation of the architectural soundness of different solutions.

Through this debate process, agents discover that the \texttt{contribute\_\allowbreak to\_\allowbreak class} approach provides superior design properties: it prevents method overwriting at the source rather than attempting runtime workarounds, requires minimal code changes at lines 765-767 in \texttt{django/\allowbreak db/\allowbreak models/\allowbreak fields/\allowbreak \_\_init\_\_.\allowbreak py} with a simple existence check using \texttt{if not hasattr(cls,\allowbreak method\_\allowbreak name)}, and maintains backward compatibility while enabling user method preservation. This debate dynamic creates productive analytical tension where agents must defend their approaches against alternatives~\cite{liu2025truth,zhang2025gdesigner}, transforming issue resolution from individual exploration to structured multi-perspective reasoning that enables precise architectural understanding and modification plans that individual agents cannot achieve~\cite{qian2025scaling,zhuge2024gptswarm}.

\section{Methodology}
\label{sec:approach}

\begin{figure*}[t]
    \centering
    \includegraphics[width=0.9\textwidth]{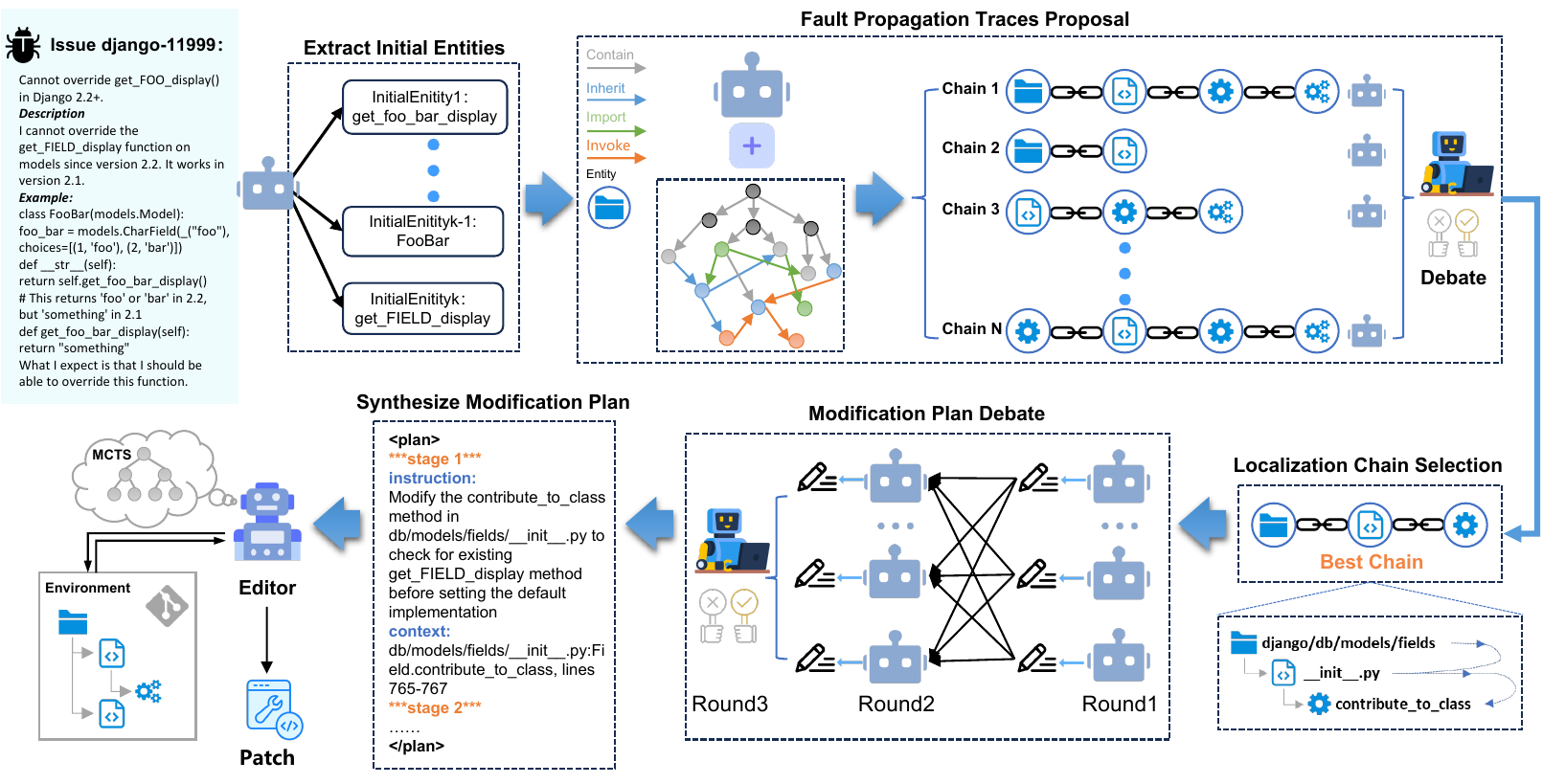}
    \vspace{-0.2cm}
    \caption{Overview of \approach framework.}
    \vspace{-0.2cm}
    \label{fig:framework}
\end{figure*}

\subsection{Problem Formulation}
\label{method:formulation}
Given an issue description $p$ and a codebase state represented as a set of code entities $V = \{v_1, \dots, v_m\}$, an agent must identify the subset of entities to be modified $V_{mod} \subset V$ through an exploration chain $\{(a_1, o_1), \dots, (a_n, o_n)\}$, where $a_t$ and $o_t$ denote the action and observation at time step $t$.
Current approaches, however, face two fundamental limitations. First, their exploration is inefficient as it overlooks structural relationships between code entities (e.g., classes, methods). Second, they struggle with modification disambiguation when multiple locations seem relevant but require different reasoning perspectives for a correct evaluation.

Our approach addresses both limitations through competitive reasoning on a code dependency graph $G = (V, E)$, where $V$ is the set of code entities and $E$ represents their structural relationships. We construct localization chains $C = (v_1, \ldots, v_k)$ composed of code entities $v_i \in V$ that trace fault propagation paths. Subsequently, we employ a multi-agent debate to resolve modification disambiguation and determine the most effective fix plan.

Our framework operates on the principle that accurate fault localization requires \textit{diverse structural viewpoints} combined with \textit{rigorous evaluation mechanisms}, as illustrated in Figure~\ref{fig:framework}. To achieve this, we employ a dual-stage competitive debate architecture. The first stage (Section~\ref{method:graph}) tackles the structural exploration problem by efficiently identifying potential fault propagation paths through graph traversal. The second stage (Section~\ref{sec:debate}) performs competitive debate on chain selection and modification disambiguation among multiple agents. Finally, the selected modification plan is integrated into an MCTS-based agentic framework for patch generation (Section~\ref{method:integration}).

\subsection{Fault Propagation Traces Proposal}
\label{method:graph}

Since issue descriptions rarely pinpoint exact modification locations, agents need to identify the fault propagation trace for issue resolution. To enable competitive fault localization, our algorithm generates multiple candidate fault propagation traces as localization proposals, followed by a competitive multi-agent debate to select the most promising trace. A fault propagation trace refers to a structured chain of code entities (e.g., classes, methods, functions, variables) that reflect how defects may propagate through the codebase.

\noindent\textbf{Dependency Graph Construction.}
The algorithm begins by building a static dependency graph $G = (V, E)$, where the node set $V$ represents code entities and the edge set $E$ captures their dependency relationships including function calls, class inheritance, module imports, and variable references. 
This graph serves as the structural backbone for tracing potential fault propagation traces, allowing agents to explore the codebase in a systematic way.

\noindent\textbf{Identifying Entry Nodes via Semantic Matching.}
Next, the algorithm identifies the top-$K$ entities $E_p = {e_1, \ldots, e_{K}}$ that are most relevant to the issue description through semantic matching. We employ a language model-based approach to extract structural identifiers that are explicitly referenced in the issue text from the dependency graph represented as an adjacency matrix and entity metadata. These entities encompass specific code-level structural elements such as class names, function names, parameter names, or error names (e.g., "UserSession", "Redis", "wholesale") that appear directly in the issue description. The extraction process prioritizes such structural identifiers while maintaining diversity through deduplication mechanisms. To ensure precision, the selection is constrained to only include entities with direct textual correspondence in the issue description, preventing the introduction of spurious entry points that could mislead subsequent chain construction.
These high-confidence entities act as entry points for subsequent chain construction. 

\noindent\textbf{Chain Construction via Graph Traversal.}
To capture diverse propagation patterns, we systematically traverse the dependency graph using a two-phase strategy for each seed entity $e_i \in E_p$: 
(1) Breadth-First Expansion: We identify top-$W$ most issue-relevant neighboring nodes based on semantic and structural relevance. We first extract the top-K entities from the issue text, which are strongly associated with the described problem. However, in many cases, the root cause of an issue is not explicitly linked to specific functions or components within the issue description. To address this, we expand the context by retrieving code snippets related to each of the top-K entities using the dependency graph. These code snippets, along with the original issue, are then fed into the LLM individually. This enriched context allows the model to identify more diverse and informative entities as potential starting points for the localization chain, effectively expanding the search space and improving localization accuracy.
(2) Depth-First Search: From each selected neighbor, we perform a depth-limited traversal (maximum depth of $L$), selecting the most promising next entity at each step. The selection is guided by a composite scoring function that considers both semantic similarity to the issue and structural importance in the dependency graph.
This process results in a total of {top-$K$ $\times$ top-$W$} localization chains, each representing a plausible fault propagation path.

These localization chains capture fault propagation patterns through dependency relationships—defects in one component affecting dependent components via call chains, inheritance hierarchies, or data flow. This structured approach efficiently identifies propagation paths that would require extensive exploration to discover through search-based methods, enabling the competitive debate process described next.

\subsection{Multi-Agent Debate}
 \label{sec:debate}

\label{method:debate}

With the localization chains identified through graph-guided analysis, we synthesize a consolidated fix plan through a competitive multi-agent debate. This requires evaluating competing architectural approaches—some chains may target core system components while others suggest more localized fixes, each with different implications for maintainability and system robustness. Our competitive debate forces agents to propose, defend, and refine their modification plans, ultimately converging on the most consolidated fix plan.

\noindent\textbf{Localization Chain Selection.}
From the {$K\times W$} candidate chains, we form a diverse set by selecting the longest chain plus the $(m-1)$ most distinct chains computed based on semantic embeddings. Multiple specialized agents then engage in competitive ranking, where each agent independently evaluates and ranks these $m$ chains based on their analytical perspective. Through this localization-level debate, agents must defend their chain preferences against alternatives, revealing structural insights that single-agent selection would miss. We select the chain with the highest aggregate vote as the optimal localization path, which serves as the foundation for the subsequent modification plan debate.

\noindent\textbf{Modification Plan Proposal.}
Based on the selected localization chain from the competitive debate, we generate a comprehensive modification plan, which systematically specifies the exact code locations requiring changes, the types of modifications needed, and their implementation priorities. 

The transformation from localization chains to modification proposals is implemented through a specialized prompt-driven analysis framework. Given a localization chain $C = \{e_1, e_2, ..., e_k\}$ and issue description $I$, each agent applies a structured analysis prompt that guides the examination of each entity $e_i$ in the chain. The prompt instructs agents to: (1) analyze code structure and functionality, (2) identify specific modification targets, (3) determine modification types (fix\_bug, add\_feature, refactor, optimize), (4) assess priority levels, and (5) provide implementation reasoning. The output is a structured JSON specification that maps each chain entity to concrete modification targets with precise location descriptions, priority rankings, and implementation strategies.

We use $N$ such agents to form a pool of $N$ diverse modification proposals. This multi-agent approach ensures comprehensive coverage of potential modification strategies as each agent contributes its unique analytical viewpoint to form a diverse pool of modification proposals.

\noindent\textbf{Competitive Strategy Refinement.}
Each agent reviews all proposals from the independent analysis phase and engages in structured argumentation to defend their approach while critically evaluating alternatives. This competitive refinement phase addresses the limitation of independent analysis by forcing agents to explicitly justify their reasoning against competing perspectives, revealing hidden assumptions and identifying potential weaknesses in initial proposals. Agents generate refined modification plans that incorporate insights from cross-agent critique while maintaining their specialized analytical focus, driving deeper understanding of the fault localization.

\noindent\textbf{Synthesize modification plan.}
Based on the refined proposals from the competitive refinement phase, a discriminator agent synthesizes insights from all refined proposals to produce a coherent, actionable modification plan with prioritized steps and rationale. This final selection phase is essential because competitive refinement may produce multiple valid but incompatible strategies that require unified resolution for practical implementation. The discriminator evaluates trade-offs between competing architectural approaches, considers implementation complexity and risk factors, and generates a structured plan that guides downstream repair processes with both strategic direction and tactical specificity.


This competitive process addresses the limitations of agents' limited individual observation and exploration scope by leveraging diverse specialized perspectives and finding architecturally sound solutions through argumentative rigor. The debate produces a structured modification plan with strategic insights, and this high-level plan can then be leveraged to guide the subsequent repair process.

\subsection{Patch Generation}
\label{method:integration}

Based on the modification plans, the final stage generates patches and fixes the issue by employing a Monte Carlo Tree Search (MCTS) framework~\cite{antoniades2024swesearch}. The MCTS process allows the agent to systematically explore the codebase, refine its modification plans, and iteratively evaluate the impact of each action on the codebase~\cite{ma2024lingma}. 
Unlike previous approaches that begin with arbitrary exploration~\cite{antoniades2024swesearch}, our MCTS process starts from the structured modification plan generated by the competitive debate framework, enabling focused exploration with pre-specified target locations and strategic reasoning. 

The MCTS process unfolds as a search through a tree where nodes represent states of the codebase and edges represent actions (\texttt{Search} for code exploration, \texttt{Plan} for strategic reasoning, and \texttt{Edit} for code modification). The initial branches of this search tree are constructed from the steps outlined in our modification plan, ensuring the exploration is grounded in the debate's strategic insights. The agent then iteratively navigates and expands this tree. At each step, it selects an action based on a modified Upper Confidence Bound for Trees (UCT)~\cite{antoniades2024swesearch} criterion that balances exploiting known high-reward paths with exploring less-visited states. 
After an action is executed, its outcome is assessed by a value function. This function, initially informed by the rationale from our modification plan, provides not only a numerical score but also a written explanation of the decision's quality. After each edit action, the agent can also execute existing tests and create new ones to better evaluate the current state. This qualitative feedback is then propagated back up the search tree, refining the agent's value estimates and guiding future decisions toward a successful resolution. 
This process continues until the agent reaches a satisfactory resolution of the issue or reaches a predefined exploration depth $D_{max}$.



\section{Experimental Setup}

\subsection{Research Questions}

\noindent\textbf{RQ1:} How effective is \approach in terms of repository-level issue resolution?

\noindent\textbf{RQ2:} How does each component of \approach contribute to its overall performance?


\noindent\textbf{RQ3:} How is the fault localization performance of \approach compared to other baselines?

\noindent\textbf{RQ4:} How do the depth of chains in the debate influence the performance?

\subsection{Datasets}
We evaluate \approach on the SWE-Bench-Verified dataset~\cite{openai2024introducing}, which contains 500 verified issues from SWE-bench~\cite{jimenez2024swebench}. We also evaluated on SWE-bench-Lite~\cite{jimenez2024swebench}, which contains 300 carefully selected tasks.

\subsection{Baselines}
We compare \approach with the following baselines on issue resolution:
\begin{itemize}[leftmargin=0.5cm]
\item \textbf{Agentless}~\cite{xia2024agentless}: A non-agentic pipeline that breaks down the repair process into different phases of localization, repair, and patch validation.
\item \textbf{AutoCodeRover}~\cite{zhang2024autocoderover}: A software engineering-oriented approach that combines LLMs with sophisticated code search capabilities.
\item \textbf{SWE-Agent}~\cite{yang2024sweagenta}: A custom agent-computer interface enabling LM agents to interact with repository environments through defined actions.
\item \textbf{SWE-Search}~\cite{antoniades2024swesearch}: A repository issue resolution agent that uses Monte Carlo Tree Search (MCTS) to explore the solution space.
\item \textbf{SWESynInfer}~\cite{ma2024lingma}: An open-source LLM series trained with development-process-centric data, simulating repository analysis, fault localization, and patch generation via a three-stage Chain-of-Thought workflow.
\item \textbf{OpenHands}~\cite{wang2024openhands}: An open-source platform for building general-purpose AI agents that solve software and web tasks through code, terminal, and browser interaction.
\end{itemize}

Additionally, we select the following baselines to compare the performance on fault localization:
\begin{itemize}[leftmargin=0.5cm]
\item \textbf{CodeActAgent}~\cite{wang2024openhands}: An agent that interact with environments through executing file system search commands to locate faults.
\item \textbf{LocAgent}~\cite{chen2025locagent}:  A graph-guided LLM-agent framework designed to enhance code localization through powerful multi-hop reasoning.
\item \textbf{KGComposs}~\cite{yang2025enhancing}: A framwework ridges semantic gaps in repository-level repair by constructing a repository-aware knowledge graph and leveraging path-guided reasoning to enhance LLM-based patch generation.
\end{itemize}


\subsection{Metrics}
We employ the following metrics to evaluate the performance of \approach:
\begin{itemize}[leftmargin=0.5cm]
    \item \textbf{Pass@1:} The percentage of issues that are resolved successfully within the first attempt, following the evaluation protocol established by~\cite{antoniades2024swesearch,yang2024sweagenta}. This metric directly measures the framework's ability to generate correct patches without requiring multiple iterations, representing the most practical scenario for real-world deployment.
    \item \textbf{Acc@1 (File):} The localization accuracy at top-1 predictions at file level, where a localization is considered successful only when all required modification points are included within the top-1 predicted locations~\cite{xia2024agentless,chen2025locagent}. This metric evaluates the model's capacity to precisely and comprehensively identify all code regions that require modification, providing a fine-grained assessment of fault localization performance prior to the patch generation stage.
\end{itemize}

\subsection{Implementation Details}

We implement \approach by extending the SWE-Search~\cite{antoniades2024swesearch} framework with our graph-based localization and multi-agent debate components. Due to unsuccessful testbed setup, we did not utilize it in our experiments. The code dependency graph is constructed using static analysis tools\footnote{\url{https://github.com/python/cpython/blob/3.13/Lib/ast.py}}, and the multi-agent debate employs offical DeepSeek-V3-0324~\cite{deepseek-ai2025deepseekv3} with different system prompts to simulate diverse reasoning perspectives. For the graph traversal parameters, we set $K=5$ for the number of entry points, $W=4$ for breadth-first expansion width, and $L=5$ for maximum chain length. The multi-agent debate involves $m=6$ chains for competitive ranking and $N=5$ specialized agents in the competitive refinement process. These parameters are set based on the a held out set in the full SWE-Bench dataset~\cite{jimenez2024swebench}. For baseline methods, we directly use the reported results either from the official leaderboard~\cite{jimenez2024swebench} or from the official paper or repository. For experiments on DeepSeek-V3-0324, we reproduce the results on representative baselines with their official repositories.


\section{Results}

We present experimental results addressing each research question, examining the effectiveness of \approach across multiple dimensions and providing detailed analysis of component contributions.

\subsection{RQ1: Effectiveness on Issue Resolution}

Table~\ref{tab:main} presents the main experimental results comparing \approach with state-of-the-art baselines on the SWE-Bench-Verified dataset. We observe that \approach is able to solve 207 out of 500 problems, achieving 41.4\% success rate. While individual baseline approaches show varying performance across different language models, \approach demonstrates consistent superiority over existing methods. Specifically, when comparing with methods using the same DeepSeek-V3-0324 model, \approach achieves significant improvements: 6.0\% over SWE-Search, improving from 35.4\% to 41.4\%, and 2.6\% over SWE-Agent, improving from 38.8\% to 41.4\%. 


\begin{table}[h]
    \centering
    \caption{Main effectiveness results on SWE-Bench-Verified.}
    \label{tab:main}
        \begin{tabular}{@{}llc@{}}
        \toprule
        \textbf{Method} & \quad\quad\quad\textbf{Model} & \textbf{Pass@1}  \\
        \midrule
        SWE-Agent & \includegraphics[height=1em]{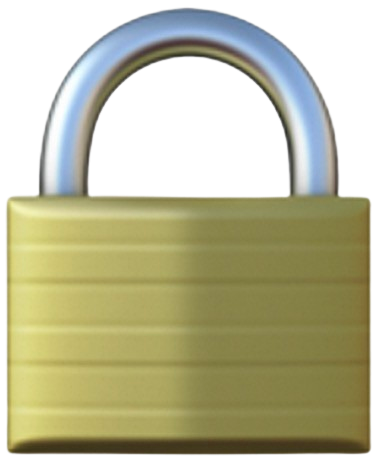} GPT-4o (2024-05-13) & 23.0\% \\
                 & \includegraphics[height=1em]{figures/lock-removebg-preview.png} Claude-3.5 Sonnet & 33.6\% \\
         & \includegraphics[height=1em]{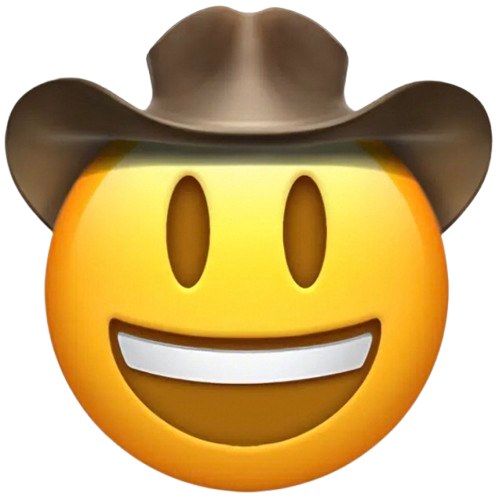} DeepSeek-V3-0324 & 38.8\% \\
        SWE-Search & \includegraphics[height=1em]{figures/smile-removebg-preview.png} DeepSeek-V3-0324 & 35.4\% \\
        Moatless Tools & \includegraphics[height=1em]{figures/smile-removebg-preview.png} DeepSeek-V3-0324 & 34.6\% \\
        Agentless & \includegraphics[height=1em]{figures/lock-removebg-preview.png} GPT-4o (2024-05-13) & 36.2\% \\
         & \includegraphics[height=1em]{figures/smile-removebg-preview.png} DeepSeek-V3-0324 & 36.6\% \\
        AutoCodeRover & \includegraphics[height=1em]{figures/lock-removebg-preview.png} GPT-4o (2024-05-13) & 38.4\% \\
        CodeAct & \includegraphics[height=1em]{figures/lock-removebg-preview.png} GPT-4o (2024-05-13) & 30.0\% \\
        SWESynInfer & \includegraphics[height=1em]{figures/lock-removebg-preview.png} Claude-3.5 Sonnet & 35.4\% \\
             & \includegraphics[height=1em]{figures/lock-removebg-preview.png} GPT-4o (2024-05-13) & 31.8\% \\
             & \includegraphics[height=1em]{figures/smile-removebg-preview.png} Lingma SWE-GPT 72B & 30.2\% \\
        OpenHands & \includegraphics[height=1em]{figures/smile-removebg-preview.png} DeepSeek-V3-0324 & 38.8\% \\
        \midrule
        \approach & \includegraphics[height=1em]{figures/smile-removebg-preview.png} DeepSeek-V3-0324 & \textbf{41.4\%} \\
        \bottomrule
        \end{tabular}
\end{table}

It is important to note that \approach outperforms even the strongest baseline configurations, including OpenHands with DeepSeek-V3-0324 and SWE-Agent with DeepSeek-V3-0324, both achieving 38.8\%. This demonstrates that our competitive multi-agent debate framework provides substantial benefits beyond what can be achieved through model selection alone. 

\begin{finding}
    \textbf{Finding 1:} \approach achieves 41.4\% Pass@1 on issue resolution, representing a 2.6 percentage point improvement over the strongest baseline using the same model, demonstrating the effectiveness of competitive multi-agent debate for repository-level issue resolution.
\end{finding}

\subsection{RQ2: Ablation Study}

To understand the contribution of each component in \approach, we conduct comprehensive ablation studies by systematically removing key components and measuring performance degradation on the SWE-Bench-Verified dataset. Table~\ref{tab:ablation} shows the results of this analysis, revealing distinct contributions from different architectural elements.

\begin{table}[h]
    \centering
    \caption{Ablation study results showing the contribution of different components.}
    \label{tab:ablation}
    \begin{tabular}{@{}lcc@{}}
    \toprule
    \textbf{Method} & \textbf{Pass@1} & \textbf{$\Delta$} \\
    \midrule
    \textbf{\approach} & \textbf{41.4\%} & - \\
    \quad w/o Multiple Chain Generation & 31.4\% & \textcolor{red}{-10.0\%} \\
    \quad w/o Multi-Agent Debate & 37.2\% & \textcolor{red}{-4.2\%} \\
    \quad w/o Edit plan & 35.4\% & \textcolor{red}{-6.0\%} \\
    \bottomrule
    \end{tabular}
\end{table}

We observe that removing the multiple chain generation component causes the most significant performance drop, with the method achieving only 31.4\% Pass@1, representing a 10.0 percentage point degradation. This suggests that exploring diverse fault propagation paths through graph traversal significantly improves localization accuracy. When this component is removed, the method must rely on single-path exploration, which frequently misses critical dependency relationships that span multiple files or modules.

Similarly, removing the edit plan component results in a 6.0 percentage point performance drop, declining from 41.4\% to 35.4\%. This demonstrates that the structured modification plans generated through competitive debate are essential for guiding the downstream patch generation process. Without these plans, the MCTS-based editing agent lacks strategic direction, leading to suboptimal exploration patterns and reduced fix accuracy.

The multi-agent debate component also plays a critical role. Its removal leads to a 4.2 percentage point drop, reducing performance to 37.2\%. This highlights the importance of competitive reasoning in resolving modification disambiguation. Without structured debate, the system relies on individual agent exploration, which often gets stuck in local solutions when multiple plausible fix locations exist. Our experiments reveal that when the localization chain contains numerous candidate files, the editing process becomes inefficient, with agents spending excessive exploration time without converging on optimal solutions.

\begin{finding}
    \textbf{Finding 2:} Multiple chain generation provides the largest contribution to performance (+10.0\%), followed by edit plan generation (+6.0\%) and Multi-agent debate (+4.2\%), demonstrating that each component addresses distinct limitations in repository-level issue resolution.
\end{finding}

\subsection{RQ3: Localization Performance Comparison}

\begin{table}[h]
    \centering
    \caption{Localization Performance on SWE-Bench-lite.} 
    \label{tab:efficiency_file}
    \begin{tabular}{@{}llc@{}}
    \toprule
    \textbf{Method} & \textbf{Model} & \textbf{Acc@1 (File)} \\
    \midrule
    Agentless & \includegraphics[height=1em]{figures/lock-removebg-preview.png} GPT-4o (2024-05-13) &  67.15\\
           & \includegraphics[height=1em]{figures/lock-removebg-preview.png} Claude-3.5 Sonnet & 72.63 \\
    SWE-Agent & \includegraphics[height=1em]{figures/lock-removebg-preview.png} GPT-4o (2024-05-13) & 57.30 \\
               & \includegraphics[height=1em]{figures/lock-removebg-preview.png} Claude-3.5 Sonnet & 77.37 \\
               & \includegraphics[height=1em]{figures/smile-removebg-preview.png} DeepSeek-V3-0324 & 67.00 \\
    SWE-Search & \includegraphics[height=1em]{figures/lock-removebg-preview.png} GPT-4o (2024-05-13) & 73.36 \\
               & \includegraphics[height=1em]{figures/lock-removebg-preview.png} Claude-3.5 Sonnet & 72.63 \\
    CodeActAgent & \includegraphics[height=1em]{figures/lock-removebg-preview.png} GPT-4o (2024-05-13) & 60.95 \\
              & \includegraphics[height=1em]{figures/lock-removebg-preview.png} Claude-3.5 Sonnet & 76.28\\
    LocAgent & \includegraphics[height=1em]{figures/smile-removebg-preview.png} Qwen2.5-7B (FT) & 70.80 \\
             & \includegraphics[height=1em]{figures/smile-removebg-preview.png} Qwen2.5-32B (FT) & 75.91 \\
             & \includegraphics[height=1em]{figures/lock-removebg-preview.png} Claude-3.5 Sonnet & 77.74\\
    KGCompass & \includegraphics[height=1em]{figures/lock-removebg-preview.png} Claude-3.5 Sonnet & 76.67 \\
    \midrule
    \textbf\approach & \includegraphics[height=1em]{figures/smile-removebg-preview.png} DeepSeek-V3-0324 & \textbf{81.67} \textcolor{Green}{(+3.93)}\\
    \bottomrule
    \end{tabular}
\end{table}

Table~\ref{tab:efficiency_file} shows the localization performance comparison across different methods on the SWE-Bench-Lite dataset, which we adopt to facilitate direct comparison with existing approaches~\cite{xia2024agentless,yang2024sweagenta,antoniades2024swesearch,lv2024codeact,chen2025locagent}, referecing the results from LocAgent~\cite{chen2025locagent}. \approach achieves 81.67\% file-level localization accuracy, significantly outperforming all baseline methods. When comparing with methods using the same DeepSeek-V3-0324 model, \approach demonstrates substantial improvements: 14.67\% over SWE-Agent, improving from 67.00\% to 81.67\%. This represents an 3.93 percentage point improvement over the strongest baseline across all model configurations, surpassing LocAgent with Claude-3.5 Sonnet which achieves 77.74\%.

The improvement in localization accuracy is largely attributed to our graph-guided approach for constructing multiple fault propagation traces. By systematically exploring code dependency relationships and generating diverse candidate chains, \approach captures structural patterns that single-pass exploration methods frequently miss. The dramatic improvement over SWE-Agent using the same model—from 67.00\% to 81.67\%—particularly highlights the effectiveness of our structured reasoning approach compared to traditional search-based methods. Compared to baseline approaches that perform localization once, our method aggregates multiple potential paths, significantly increasing the likelihood that at least one trace contains the correct fix location.

The results demonstrate that our architectural innovations provide benefits that transcend model capabilities. While methods using stronger language models like Claude-3.5 Sonnet generally achieve better localization performance than those using GPT-4o, \approach with DeepSeek-V3-0324 surpasses even the best Claude-3.5 Sonnet results. More importantly, the consistent superiority over other methods using the identical DeepSeek-V3-0324 model confirms that performance gains stem from enhanced reasoning frameworks rather than model sophistication alone.

Furthermore, the gap between our localization performance and that of specialized localization methods like LocAgent demonstrates substantial improvement, with \approach achieving 81.67\% compared to LocAgent's 77.74\%. This shows that competitive multi-agent debate can enhance even domain-specific approaches. The reason is that our debate process forces agents to systematically evaluate competing localization hypotheses, preventing premature convergence on suboptimal solutions.

\begin{finding}
    \textbf{Finding 3:} \approach achieves 81.67\% file-level localization accuracy, surpassing the strongest baseline by 3.93\%, demonstrating that graph-guided fault propagation traces combined with competitive debate enable more accurate fault localization than individual exploration approaches.
\end{finding}

\subsection{RQ4: Impact of the Chain Depth.}



To investigate the optimal configuration for multi-agent reasoning in software fault localization, we study the impact of chain depth on localization performance. To balance computational efficiency with representative evaluation, we constructed a dataset of 75 instances, termed SWE-Bench-Verified-S, consisting of 50 samples from SWE-Bench-verified-mini~\footnote{\url{https://huggingface.co/datasets/MariusHobbhahn/swe-bench-verified-mini}} and 25 additional instances randomly selected in SWE-Bench-Verified.

\begin{figure}[H]
    \centering
    \includegraphics[width=0.7\linewidth]{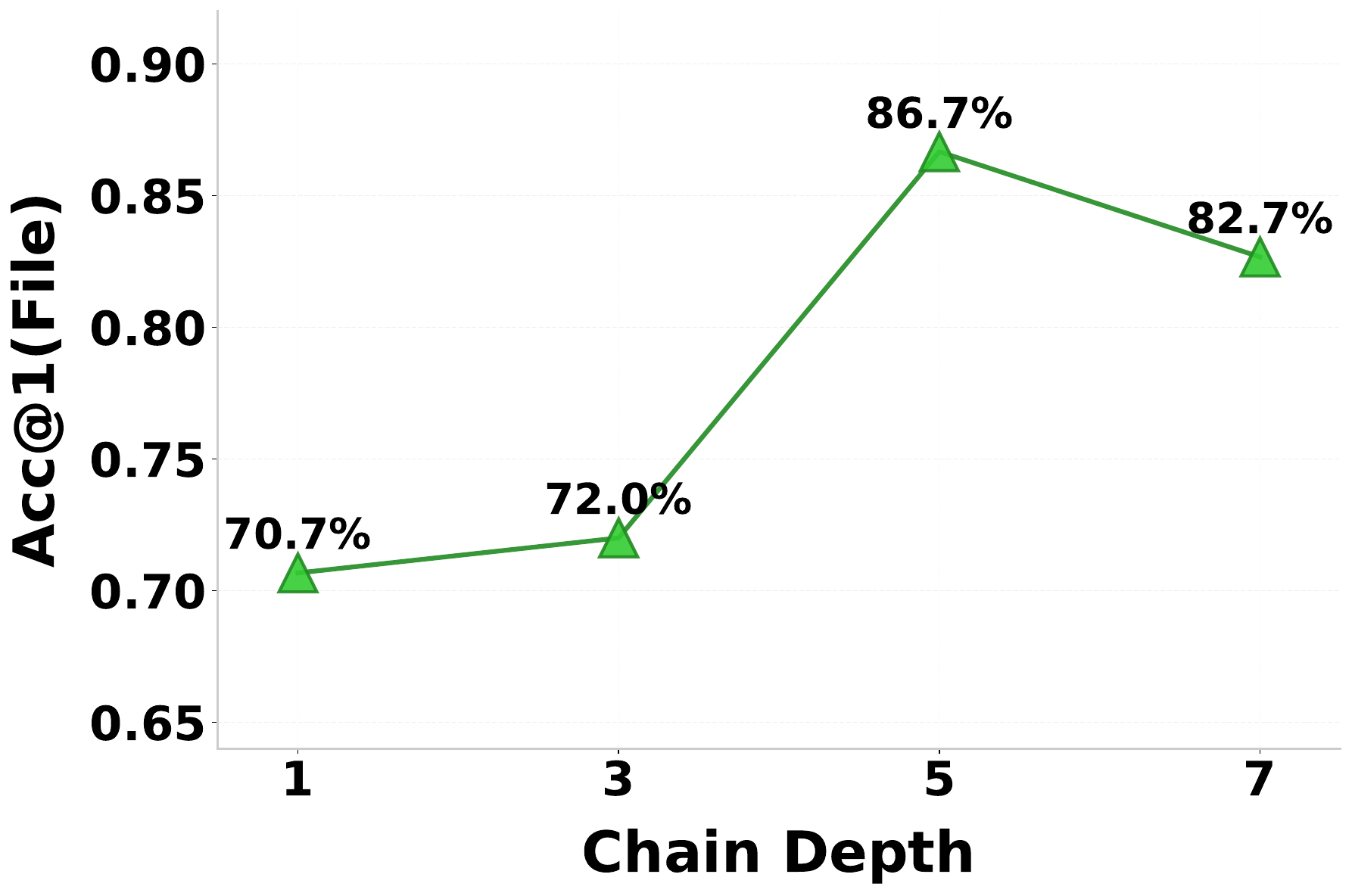}
    \vspace{-0.2cm}
    \caption{Impact of the Chain Depth.}
    \vspace{-0.2cm}
    \label{fig:rq4}  
\end{figure}

Figure~\ref{fig:rq4} shows the impact of varying chain depth in the competitive debate process on this dataset. We observe that increasing the chain depth from 1 to 5 consistently boosts file-level localization accuracy, reaching a peak Acc@1(File) of 86.7\%. This suggests that deeper reasoning chains enable more effective exploration of the code graph, capturing complex dependencies and contextual signals that shallow chains often miss. These results highlight the benefits of multi-step reasoning in guiding the model toward more informed and accurate localization decisions.

However, we also find that increasing the chain depth beyond 5 leads to diminishing returns and even slight performance degradation. This suggests a trade-off between reasoning depth and relevance. Longer chains are more likely to include information unrelated to resolving the issue, which can distract the model and reduce its ability to judge which chain is most likely to lead to a correct fix. As a result, the debate process becomes less focused, making it harder to converge on accurate localization decisions.

\begin{finding}
    \textbf{Finding 4:} A chain depth of 5 achieves the best trade-off between reasoning depth and relevance, yielding the highest localization accuracy. Further increases introduce distracting information that hinders effective decision-making during the debate.
\end{finding}

\begin{figure*}[t]
    \centering
    \includegraphics[width=0.8\textwidth]{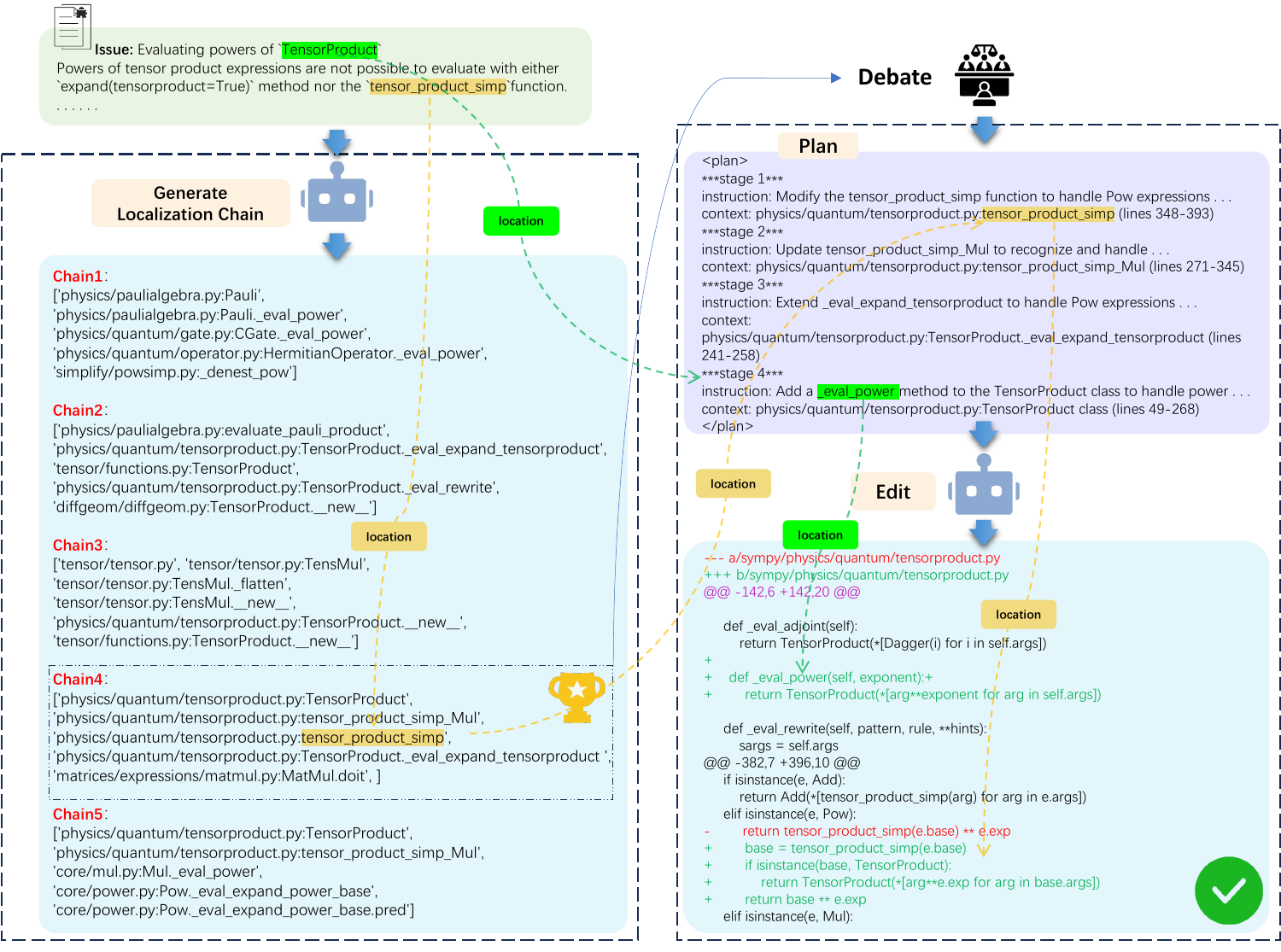}
    \vspace{-0.2cm}
    \caption{Case study of \approach with instance \textit{sympy-13974}.}
    \vspace{-0.2cm}
    \label{fig:case}
\end{figure*}

\subsection{Case Study}

To further verify the effectiveness of SWE-debate in actual use, we analyzed a case in SWE-bench. In this case, we investigate an issue in \texttt{SymPy} related to the incorrect evaluation of powers applied to \texttt{TensorProduct} expressions. Specifically, expressions like \texttt{TensorProduct(1,\allowbreak Pauli(3))\allowbreak *\allowbreak TensorProduct(1,\allowbreak Pauli(3))} fail to simplify into \texttt{TensorProduct(1,\allowbreak 1)} or \texttt{1}, even though \texttt{Pauli(3)**2\allowbreak =\allowbreak 1}. This failure arises because neither the \texttt{expand(tensorproduct\allowbreak =\allowbreak True)} method nor the \texttt{tensor\_\allowbreak product\_\allowbreak simp} function are equipped to handle exponentiation of \texttt{TensorProduct} objects. The issue is non-trivial as it requires coordinated reasoning over symbolic power expressions and tensor algebra simplification.

Our method first generates  localization chains based on graph-based reasoning over symbolic dependencies, and then applies a Multi-Agent Debate process to guide the repair. In the first stage, the agent constructs multiple candidate localization chains. Among the generated candidates, Chain 2 demonstrates high accuracy by precisely capturing all necessary modules involved in the failure, such as \texttt{tensor\_\allowbreak product\_\allowbreak simp\_\allowbreak Mul}, \texttt{tensor\_\allowbreak product\_\allowbreak simp}, and \texttt{TensorProduct.\allowbreak eval\_\allowbreak expand\_\allowbreak tensorproduct}. This chain provides a comprehensive view of the symbolic rewriting pipeline for tensor expressions, effectively guiding the system to the correct set of files and even narrowing down the specific functions requiring modification.

Based on this chain, our debate process formulates a multi-stage plan. It proposes a four-step approach: first, modifying the simplification logic in \texttt{tensor\_\allowbreak product\_\allowbreak simp}; second, updating \texttt{tensor\_\allowbreak product\_\allowbreak simp\_\allowbreak Mul} to support powers; third, extending \texttt{eval\_\allowbreak expand\_\allowbreak tensorproduct} for power handling; and fourth, adding an \texttt{\_eval\_\allowbreak power} method in the \texttt{TensorProduct} class. The edit derived from this plan introduces a recursive rule that distributes the exponent over the arguments of a \texttt{TensorProduct}, enabling correct evaluation of symbolic powers. Notably, all modified functions were covered by the selected localization chain, demonstrating the accuracy and completeness of our code navigation strategy.

This case highlights the strength of our method in both locating and resolving issues. The localization chain effectively surfaces the relevant symbolic manipulation points, while the structured planning and debate framework facilitates coordinated edits across multiple modules. Together, they contribute to generating a valid and verifiable patch that resolves the issue as expected.

\section{Discussion}

\subsection{Strengths}

\approach demonstrates three key advantages over existing approaches.
First, the graph-based localization significantly improves fault localization accuracy. By building dependency graphs from code structure and generating multiple fault propagation traces, our method achieves 81.67\% file-level accuracy. When comparing with methods using identical models, \approach demonstrates substantial improvements: 14.67\% over SWE-Agent with DeepSeek-V3-0324 and 8.31\% over SWE-Search with GPT-4o , representing an 3.93 percentage point improvement over the strongest baseline across all configurations. The graph traversal strategy effectively captures structural relationships that single-agent individual exploration often misses, particularly for multi-file issues where traditional methods frequently fail due to insufficient architectural understanding. This superior localization performance directly translates to better issue resolution, with \approach achieving 41.4\% Pass@1 compared to baseline methods ranging from 23.0\% to 38.8\%.

Second, the competitive multi-agent debate resolves modification ambiguity more effectively than individual agent reasoning. Our ablation study shows that removing the debate component causes a 4.2 percentage point drop in resolution rate. The three-round debate process—independent analysis, competitive refinement, and final selection—enables agents with different specializations to systematically evaluate competing fix strategies. This structured competitive approach achieves a 2.6 percentage point improvement over the strongest baseline using the same model, demonstrating that architectural innovations can transcend model capabilities when multiple code locations appear relevant but require different architectural considerations for correct resolution.

Third, the framework integrates seamlessly with existing issue resolution systems without requiring major modifications. Our approach can improve the localization modules in frameworks like SWE-Search and Agentless, providing better starting points for downstream repair while maintaining compatibility with their existing architectures. This plug-and-play design enables practical adoption in real software engineering workflows.

\subsection{Limitations and Future Work}

Despite the promising results, \approach has several limitations that suggest directions for future work. The graph construction process can be computationally expensive for large codebases, limiting scalability. Future work chould explore more efficient graph construction algorithms and incremental analysis techniques to handle enterprise-scale repositories. Additionally, the current static analysis approach may miss dynamic relationships and runtime behaviors that could improve localization accuracy for certain types of issues. 
Our multi-agent debate currently relies on a single model with different prompts to simulate diverse reasoning perspectives, which may not fully capture the breadth of real-world developer reasoning styles. While our specialized prompts enforce distinct analytical viewpoints and our ablation study confirms significant performance gains from the debate mechanism, integrating multiple heterogeneous models or incorporating domain-specific knowledge bases could further enhance the diversity and quality of the debate process. Future work could explore how different foundation models with varying reasoning capabilities can be orchestrated within the competitive debate framework to achieve even greater analytical diversity.
The current batch processing approach limits integration with real-time development workflows. Future work could investigate lightweight continuous analysis modes and tighter integration with development environments to provide immediate issue resolution assistance during coding.

\section{Threats to Validity}

\textbf{Internal.} The primary internal threat stems from potential data contamination, as the pre-training corpus of DeepSeek-V3-0324 may contain repositories from SWE-Bench. To mitigate this concern, we emphasize that our evaluation focuses on reasoning processes rather than memorized solutions. Our method generates fault propagation traces through systematic graph traversal and structured multi-agent debate, relying on analytical reasoning rather than direct code recall. The substantial improvements over baseline methods using identical models provide evidence that performance gains derive from enhanced reasoning capabilities rather than memorization effects. Future work will include evaluation on contamination-free datasets to further validate these findings.

A second internal threat arises from experimental scope limitations imposed by time and budget constraints. Our evaluation is restricted to the open-source DeepSeek-V3-0324 model and a subset of the SWE-Bench-Verified dataset. While our results demonstrate that \approach outperforms numerous methods across both identical and different foundation models, this constraint limits our ability to comprehensively validate the generalizability of our competitive debate framework across diverse language model architectures. Future work will expand evaluation to encompass a broader range of foundation models and larger datasets to establish more comprehensive performance benchmarks.


\textbf{External.} The main external threat comes from evaluation on a single dataset SWE-Bench-Verified limited to Python repositories, which may not generalize to other programming languages or software domains. To address this, our key components—dependency graph construction, semantic matching, and debate frameworks—are designed to be language-agnostic, focusing on structural reasoning rather than language-specific patterns. And we will evaluate our method on more diverse datasets like Multi-SWE-Bench~\cite{zan2025multiswebench} in the future.


\section{Related Work}

\subsection{Fault Localization}
Traditional fault localization techniques, including spectrum-based fault localization (SBFL)~\cite{jones2005empirical}, mutation-based fault localization (MBFL)~\cite{papadakis2015metallaxisfl}, and learning-based methods~\cite{sohn2017fluccs, li2019deepfl, meng2022improving}, mainly relied on test execution data and program analysis to identify buggy code regions. However, these techniques face fundamental limitations when applied to repository-level issue resolution: they require comprehensive test suites to trigger fault patterns~\cite{mundler2024swtbench, wang2024aegis}, struggle with complex dependency relationships across multiple files, and cannot effectively bridge the semantic gap between natural language issue descriptions and code structures.

Recent LLM-based approaches have advanced fault localization through sophisticated code understanding and repository navigation capabilities. Methods like RCAgent~\cite{wang2024rcagent} integrate multiple analysis tools for decision support, and AgentFL~\cite{qin2025agentfl} scales fault localization through multi-agent collaboration with static analysis tools~\cite{yang2024large,wu2023large,qin2025agentfl,wang2024rcagent}. LocAgent~\cite{chen2025locagent} leverages graph-based representations to enable multi-hop reasoning across code dependencies, while OrcaLoca~\cite{yu2025orcalocab} improves localization accuracy through priority-based scheduling and distance-aware context pruning. And CoSIL~\cite{jiang2025cosil} reduces search space using module call graphs with iterative context-aware exploration. However, these methods remain fundamentally limited by single-agent reasoning perspectives and struggle with modification disambiguation scenarios where multiple locations match issue descriptions but require different architectural viewpoints for correct evaluation. Our work addresses this limitation by introducing competitive multi-agent debate that systematically evaluates competing localization hypotheses.

\subsection{Repository-Level Issue Resolution}

Automated repository issue resolution has evolved through two main paradigms-agent-based and pipeline-based. Agent-based systems model software engineering as sequential decision-making, where language models interact with code environments through structured action spaces. SWE-Agent~\cite{yang2024sweagenta} established foundational principles for agent-environment interaction, AutoCodeRover~\cite{zhang2024autocoderover} focused on search-based localization, SWE-Search~\cite{antoniades2024swesearch} introduced Monte Carlo Tree Search for systematic exploration, and CodeR~\cite{chen2024coder} explored collaborative multi-agent architectures with pre-defined task graphs. Pipeline-based approaches break down issue resolution into specialized computational workflows. Agentless~\cite{xia2024agentless} pioneered this paradigm by separating localization, repair, and validation into targeted stages, while CodeMonkeys~\cite{ehrlich2025codemonkeys} investigated iterative refinement through test-time computation scaling. Recent advances include long-context models with appropriate prompting~\cite{jiang2025putting}, training-based approaches for specialized model fine-tuning~\cite{pham2025swesynth, pan2024training, yang2025swesmith}, and RepoUnderstander~\cite{ma2024how} which constructs repository knowledge graphs for enhanced whole-repository understanding.

However, existing methodologies face a fundamental limitation stemming from limited observation scope. They often get stuck in local solutions and fail to resolve ambiguities when multiple code locations appear plausible, as they lack the diverse analytical perspectives needed to systematically evaluate competing modification plans~\cite{zhang2025swebench, antoniades2024swesearch, cuadron2025danger}. Our approach targets this localization bottleneck by providing more accurate fault localization through competitive multi-agent analysis, enabling seamless integration with current issue resolution systems while improving their overall issue resolution rates.

\subsection{LLM Multi-Agent Systems}

Multi-agent systems have emerged as a promising approach for complex problem-solving by leveraging diverse specialized perspectives and collaborative reasoning. In software engineering contexts, these systems have shown success across various tasks including code generation~\cite{huang2023agentcoder, wu2023autogen}, automated testing and debugging~\cite{hong2023metagpt,shi2024code}. Current multi-agent architectures mainly use collaborative paradigms that emphasize consensus-building and information sharing, with scaling achieved through either cognitive enhancement of individual agents or population scaling through large agent collectives~\cite{zhuge2024gptswarm, qian2024scaling}.

Multi-agent debate systems represent a particularly relevant approach for decision-making scenarios requiring systematic evaluation of competing alternatives. Existing debate frameworks typically follow collaborative models where agents engage in structured argumentation to reach consensus through iterative refinement~\cite{chan2023chateval, du2024improving}. However, these collaborative approaches face critical limitations in technical domains: agents often suffer from thought degeneration and resist modification despite potentially incorrect stances. Recent work has attempted to address these issues through role assignment strategies and agreement modulation techniques~\cite{zhang2025gdesigner,liu2025truth}, but these approaches still maintain collaborative consensus-seeking paradigms that may not generate sufficient analytical pressure for complex architectural decision-making. In contrast, our work introduces a competitive debate framework for software fault localization that creates analytical tension by forcing agents to rigorously defend their localization hypotheses against competing proposals. Our structured, multi-round debate, combined with graph-based dependency analysis, is designed to excel at tasks requiring precise disambiguation and strategic architectural reasoning, overcoming the limitations of purely collaborative systems.

\section{Conclusion}

In this paper, we presented \approach, a competitive multi-agent debate framework that addresses the agents' limited observation scope problem in repository-level issue resolution. Our approach combines graph-based fault propagation trace generation with structured multi-agent debates to systematically evaluate competing localization hypotheses, overcoming the perspective limitations of single-agent methods. Experimental evaluation on SWE-Bench-Verified shows that \approach achieves 6.7\% improvement in issue resolution rate over state-of-the-art baselines. The framework also demonstrates 5.1\% improvement in fault localization accuracy, with potential to enhance the fault localization phases of other automated issue resolution methods. 


\section*{Acknowledgment}
This research is funded by the National Key Research and Development Program of China (Grant No. 2023YFB4503802) and the Natural Science Foundation of Shanghai (Grant No. 25ZR1401175).

\bibliographystyle{ACM-Reference-Format}
\bibliography{ref}

\appendix
\section{SWE-Bench-Verified-S}
SWE-Bench-verified-mini\footnote{\url{https://huggingface.co/datasets/MariusHobbhahn/swe-bench-verified-mini}} is a subset of SWE-Bench-Verified, containing 50 instead of 500 datapoints, requiring 5GB instead of 130GB of storage, while maintaining a similar distribution of performance, test pass rates, and task difficulty as the original dataset.
Building on SWE-Bench-verified-mini, we augment it with 25 additional instances to better approximate the distribution and performance characteristics of the full dataset, resulting in our constructed benchmark, SWE-Bench-Verified-S.

\begin{table}[h]
\small
\centering
\caption{Instance Id in SWE-Bench-Verified-S}
\label{tab:swe_verified_s}
\begin{tabularx}{\linewidth}{>{\centering\arraybackslash}X >{\centering\arraybackslash}X}
\toprule
django\_\_django-11790 & django\_\_django-11815 \\
django\_\_django-11848 & django\_\_django-11880 \\
django\_\_django-11885 & django\_\_django-11951 \\
django\_\_django-11964 & django\_\_django-11999 \\
django\_\_django-12039 & django\_\_django-12050 \\
django\_\_django-12143 & django\_\_django-12155 \\
django\_\_django-12193 & django\_\_django-12209 \\
django\_\_django-12262 & django\_\_django-12273 \\
django\_\_django-12276 & django\_\_django-12304 \\
django\_\_django-12308 & django\_\_django-12325 \\
django\_\_django-12406 & django\_\_django-12708 \\
django\_\_django-12713 & django\_\_django-12774 \\
django\_\_django-9296  & sympy\_\_sympy-13852 \\
sympy\_\_sympy-12481  & sympy\_\_sympy-17318 \\
sympy\_\_sympy-16766  & sympy\_\_sympy-15976 \\
sympy\_\_sympy-13974  & sympy\_\_sympy-13798 \\
sympy\_\_sympy-13647  & sympy\_\_sympy-20916 \\
sympy\_\_sympy-12489  & sympy\_\_sympy-24562 \\
sympy\_\_sympy-23824  & sympy\_\_sympy-23950 \\
sympy\_\_sympy-24661  & sympy\_\_sympy-16792 \\
sympy\_\_sympy-18189  & sympy\_\_sympy-12096 \\
sympy\_\_sympy-24539  & sympy\_\_sympy-13757 \\
sympy\_\_sympy-19495  & sympy\_\_sympy-18698 \\
sympy\_\_sympy-19346  & sympy\_\_sympy-17139 \\
sympy\_\_sympy-15809  & sympy\_\_sympy-22456 \\
sphinx-doc\_\_sphinx-10323 & sphinx-doc\_\_sphinx-10435 \\
sphinx-doc\_\_sphinx-10466 & sphinx-doc\_\_sphinx-10673 \\
sphinx-doc\_\_sphinx-11510 & sphinx-doc\_\_sphinx-7590 \\
sphinx-doc\_\_sphinx-7748  & sphinx-doc\_\_sphinx-7757 \\
sphinx-doc\_\_sphinx-7985  & sphinx-doc\_\_sphinx-8035 \\
sphinx-doc\_\_sphinx-8056  & sphinx-doc\_\_sphinx-8265 \\
sphinx-doc\_\_sphinx-8269  & sphinx-doc\_\_sphinx-8475 \\
sphinx-doc\_\_sphinx-8548  & sphinx-doc\_\_sphinx-8551 \\
sphinx-doc\_\_sphinx-8638  & sphinx-doc\_\_sphinx-8721 \\
sphinx-doc\_\_sphinx-9229  & sphinx-doc\_\_sphinx-9230 \\
sphinx-doc\_\_sphinx-9281  & sphinx-doc\_\_sphinx-9320 \\
sphinx-doc\_\_sphinx-9367  & sphinx-doc\_\_sphinx-9461 \\
sphinx-doc\_\_sphinx-9698  & \\
\bottomrule
\end{tabularx}
\end{table}

\section{Hyperparameters of MCTS}
\label{appendix:mcts}

The Monte Carlo Tree Search (MCTS) algorithm used in this study employs several hyperparameters as following~\cite{jimenez2024swebench}:

\begin{table}[t]
\centering
\caption{MCTS Hyperparameters}
\label{tab:mcts-hyperparams}
\resizebox{\linewidth}{!}{%
\begin{tabular}{@{}llc@{}}
\toprule
\textbf{Hyperparameter} & \textbf{Description} & \textbf{Default} \\
\midrule
\multicolumn{3}{@{}l}{\textit{Main Search Parameters}} \\
\quad c\_param & UCT exploration parameter & 1.41 \\
\quad max\_expansions & Max children per node & 2 \\
\quad max\_iterations & Max MCTS iterations & 20 \\
\quad provide\_feedback & Enable feedback & True \\
\quad best\_first & Use best-first strategy & True \\
\quad value\_function\_temperature & Value function temperature & 0.2 \\
\quad max\_depth & Max tree depth & 20 \\
\addlinespace
\multicolumn{3}{@{}l}{\textit{UCT Score Calculation Parameters}} \\
\quad exploration\_weight & UCT exploration weight & 1.0 \\
\quad depth\_weight & Depth penalty weight & 0.8 \\
\quad depth\_bonus\_factor & Depth bonus factor & 200.0 \\
\quad high\_value\_threshold & High-value node threshold & 55.0 \\
\quad low\_value\_threshold & Low-value node threshold & 50.0 \\
\quad very\_high\_value\_threshold & Very high-value threshold & 75.0 \\
\quad high\_value\_leaf\_bonus\_constant & High-value leaf bonus & 20.0 \\
\quad high\_value\_bad\_children\_bonus\_constant & High-value bad children bonus & 20.0 \\
\quad high\_value\_child\_penalty\_constant & High-value child penalty & 5.0 \\
\addlinespace
\multicolumn{3}{@{}l}{\textit{Action Model Parameters}} \\
\quad action\_model\_temperature & Action model temperature & 0.7 \\
\addlinespace
\multicolumn{3}{@{}l}{\textit{Discriminator Parameters}} \\
\quad number\_of\_agents & Number of Discriminator Agents & 5 \\
\quad number\_of\_round & Number of debate rounds & 3 \\
\quad discriminator\_temperature & Discriminator temperature & 1 \\
\bottomrule
\end{tabular}%
}
\end{table}

\section{Prompt Templates}
In the following section, we enumerate all the prompts used throughout our entire workflow, from the initial entity extraction to the final plan generation.


\begin{tcolorbox}[promptbox={Prompt 1: INITIAL ENTITY EXTRACTION PROMPT}]
\begin{lstlisting}[basicstyle=\ttfamily\footnotesize, breaklines=true]
You are a code analysis expert. Given an issue description, your task is to identify the most relevant code entities (classes, methods, functions, variables) that are likely involved in the issue.

Important: Only extract entities that are explicitly mentioned or strongly implied by the issue description. Do not invent names that are not referenced in the text.

**Issue Description:**
{issue_description}

**Instructions:**
1. Analyze the issue description to identify:
   - **Classes**: e.g., `UserAuthenticator`, `PaymentProcessor`
   - **Methods/Functions**: e.g., `validate_credentials()`, `process_payment()`
   - **Variables/Parameters**: e.g., `user_id`, `transaction_amount`
   - **Error Types/Exceptions**: e.g., `RateLimitExceededError`, `DatabaseConnectionError`
2. **Focus on direct mentions**: Only include entities that are clearly referenced in the issue.
3. **Avoid redundancy**: If multiple terms refer to the same entity (e.g., "the payment handler" and `PaymentProcessor`), pick the most precise name.
4. **Prioritize key components**: Rank entities by how central they are to the issue.
5. **Return only names**: Do not include paths, modules, or extra descriptions.
6. **Limit to {max_entities} entities**: Select only the {max_entities} most relevant and important entities for this issue.

**Output Format:**
Return a JSON list of exactly {max_entities} entity names in order of relevance (most relevant first):
["entity_name1", "entity_name2", "entity_name3", ...]

**Examples:**

1. **Issue Description:**
    Query syntax error with condition and distinct combination
    Description:
    A Count annotation containing both a Case condition and a distinct=True param produces a query error on Django 2.2 (whatever the db backend). A space is missing at least (... COUNT(DISTINCTCASE WHEN ...).

   **Output (if max_entities=3):**
   ["Count", "DISTINCTCASE", "distinct"]

2. **Issue Description:**
   "After upgrading to v2.0, the `UserSession` class sometimes fails to store session data in Redis, causing login loops."

   **Output (if max_entities=2):**
   ["UserSession", "Redis"]

3. **Issue Description:**
   "The `calculate_discount()` function applies incorrect discounts for bulk orders when `customer_type = 'wholesale'`."

   **Output (if max_entities=3):**
   ["calculate_discount", "customer_type", "wholesale"]

Note: Return only the simple names like "__iter__", "page_range", "MyClass", "my_function", etc. Do not include file paths or full qualified names.
Return exactly {max_entities} entities, prioritizing the most important ones if there are more candidates.
\end{lstlisting}
\end{tcolorbox}

\begin{tcolorbox}[promptbox={Prompt 2: CODE SNIPPET ENTITY EXTRACTION PROMPT}]
\begin{lstlisting}[basicstyle=\ttfamily\footnotesize, breaklines=true]
Based on the following code snippets and problem statement, identify the 4 most relevant entities (files, classes, or functions) that are likely involved in solving this issue.

**Problem Statement:**
{problem_statement}

**Code Snippets:**
{code_snippets}

**Instructions:**
1. Analyze the problem statement to understand what needs to be fixed/implemented
2. Review the code snippets to identify relevant entities
3. **PRIORITIZE DIVERSITY**: Select entities from different files whenever possible to ensure comprehensive coverage
4. **BALANCE RELEVANCE AND DIVERSITY**: Choose entities that are both highly relevant to the issue AND come from different modules/files
5. Avoid selecting multiple entities from the same file unless absolutely necessary
6. Select exactly 4 entities that collectively provide the best coverage for solving the issue
7. For each entity, provide the exact entity ID in the format expected by the codebase

**Selection Strategy:**
- First priority: High relevance to the problem + Different file locations
- Second priority: High relevance to the problem (even if some files overlap)
- Ensure the selected entities represent different aspects or layers of the solution

**Output Format:**
Return a JSON list containing exactly 4 entities, each with the following format:
```json
[
    {{
        "entity_id": "file_path:QualifiedName or just file_path",
        "entity_type": "file|class|function", 
        "relevance_reason": "Brief explanation of why this entity is relevant to the issue",
        "diversity_value": "How this entity adds diversity (e.g., 'different file', 'different layer', 'different functionality')"
    }}
]
```

**Example:**
```json
[
    {{
        "entity_id": "src/models.py:UserModel",
        "entity_type": "class",
        "relevance_reason": "Contains user-related functionality mentioned in the issue",
        "diversity_value": "Model layer from different file"
    }},
    {{
        "entity_id": "src/views.py:UserView",
        "entity_type": "class", 
        "relevance_reason": "Handles user interface logic that may need modification",
        "diversity_value": "View layer from different file"
    }},
    {{
        "entity_id": "src/utils/validators.py:validate_user_input",
        "entity_type": "function",
        "relevance_reason": "Input validation logic relevant to the user issue",
        "diversity_value": "Utility function from different module"
    }},
    {{
        "entity_id": "src/config.py",
        "entity_type": "file",
        "relevance_reason": "Configuration settings that may affect user behavior",
        "diversity_value": "Configuration file from different location"
    }}
]
```

**Remember**: Maximize both relevance to the issue AND diversity across different files/modules to ensure comprehensive localization chain generation.
\end{lstlisting}
\end{tcolorbox}

\begin{tcolorbox}[promptbox={Prompt 3: NEIGHBOR PREFILTERING PROMPT}]
\begin{lstlisting}[basicstyle=\ttfamily\footnotesize, breaklines=true]
You are a code analysis expert helping to select the most relevant and diverse neighbors for exploring a dependency graph to solve a specific issue.

**Issue Description:**
{issue_description}

**Current Entity:** {current_entity}
**Current Entity Type:** {current_entity_type}
**Traversal Depth:** {depth}

**Available Neighbor Entities ({total_count} total):**
{neighbor_list}

**Your Task:**
From the {total_count} available neighbors, select up to {max_selection} most relevant and diverse entities that would be most promising to explore next.

**Selection Criteria:**
1. **Relevance to Issue**: How likely is this neighbor to contain code related to solving the issue?
2. **Diversity**: Avoid selecting too many entities from the same file or with similar names
3. **Strategic Value**: Prioritize entities that could lead to discovering the root cause or solution
4. **Entity Type Variety**: Balance between files, classes, and functions when possible

**Instructions:**
1. Analyze each neighbor entity ID to understand what it likely represents
2. Consider file paths, entity names, and types to assess relevance
3. Ensure diversity by avoiding redundant selections from the same file/module
4. Select entities that complement each other in exploring different aspects of the issue
5. Return exactly the entity IDs that should be explored further (up to {max_selection})

**Output Format:**
Return a JSON object with your selection:
```json
{{
    "selected_neighbors": [
        "neighbor_entity_id_1",
        "neighbor_entity_id_2", 
        ...
    ],
    "selection_reasoning": "Brief explanation of your selection strategy and why these neighbors were chosen",
    "diversity_considerations": "How you ensured diversity in your selection"
}}
```

Focus on strategic exploration that maximizes the chance of finding issue-relevant code while maintaining diversity.
\end{lstlisting}
\end{tcolorbox}

\begin{tcolorbox}[promptbox={Prompt 4: NODE SELECTION PROMPT}]
\begin{lstlisting}[basicstyle=\ttfamily\footnotesize, breaklines=true]
You are a code analysis expert helping to navigate a dependency graph to solve a specific issue. Given the current context and available neighboring nodes, determine which node would be most promising to explore next.

**Issue Description:**
{issue_description}

**Current Entity:** {current_entity}
**Current Entity Type:** {current_entity_type}
**Traversal Depth:** {depth}

**Available Neighbor Nodes:**
{neighbor_info}

**Context:**
- We are performing graph traversal to find code locations relevant to solving this issue
- Each neighbor represents a related code entity (file, class, or function)
- We need to select the most promising node to continue exploration

**Instructions:**
1. Analyze how each neighbor might relate to solving the issue
2. Consider the traversal depth and whether we should continue or stop
3. Evaluate which neighbor is most likely to contain relevant code for the solution
4. Return your decision on whether to continue exploration and which neighbor to select

**Output Format:**
Return a JSON object with your decision:
```json
{{
    "should_continue": true/false,
    "selected_neighbor": "neighbor_entity_id or null",
    "reasoning": "Explanation of your decision",
    "confidence": 0-100
}}
```

If should_continue is false, set selected_neighbor to null.
If should_continue is true, select the most promising neighbor_entity_id.
\end{lstlisting}
\end{tcolorbox}

\begin{tcolorbox}[promptbox={Prompt 5: CHAIN VOTING PROMPT}]
\begin{lstlisting}[basicstyle=\ttfamily\footnotesize, breaklines=true]
You are an expert software engineer tasked with identifying the optimal modification location for solving a specific software issue.

**Issue Description:**
{issue_description}

**Available Localization Chains:**
{chains_info}

**Your Task:**
Analyze each localization chain as a potential modification target and vote for the ONE chain where making changes would most likely resolve the issue described above.

**Evaluation Criteria:**
1. **Problem Location Accuracy**: Does this chain contain the actual location where the bug/issue manifests?
2. **Modification Impact**: How directly would changes to this code path affect the described problem?
3. **Code Modifiability**: Is the code in this chain well-structured and safe to modify?
4. **Solution Completeness**: Would fixing this chain likely resolve the entire issue, not just symptoms?
5. **Risk Assessment**: What are the risks of modifying this particular code path?

**Key Questions to Consider:**
- Which chain contains the root cause rather than just related functionality?
- Where would a developer most likely need to make changes to fix this specific issue?
- Which code path, when modified, would have the most direct impact on resolving the problem?
- Which chain provides the clearest entry point for implementing a fix?

**Instructions:**
1. For each chain, analyze whether modifying its code would directly address the issue
2. Consider the logical flow: which chain is most likely to contain the problematic code?
3. Evaluate implementation feasibility: which chain would be safest and most effective to modify?
4. Vote for exactly ONE chain that represents the best modification target
5. Focus on where to make changes, not just what's related to the issue

**Output Format:**
Return a JSON object with your vote:
```json
{{
    "voted_chain_id": "chain_X",
    "confidence": 85,
    "reasoning": "Detailed explanation of why this chain is the best modification target for solving the issue",
    "modification_strategy": "Brief description of what type of changes would be needed in this chain",
    "chain_analysis": {{
        "chain_1": "Assessment of this chain as a modification target",
        "chain_2": "Assessment of this chain as a modification target",
        ...
    }}
}}
```

**Example:**
```json
{{
    "voted_chain_id": "chain_2",
    "confidence": 88,
    "reasoning": "Chain 2 contains the pagination iterator __iter__ method which is where the infinite loop issue described in the problem statement actually occurs. Modifying the logic in this method to properly handle the iteration termination would directly solve the reported bug.",
    "modification_strategy": "Add proper boundary checking and iteration termination logic in the __iter__ method",
    "chain_analysis": {{
        "chain_1": "Contains utility functions but modifications here would not address the core iteration logic issue",
        "chain_2": "Contains the actual iterator implementation where the bug manifests - ideal modification target",
        "chain_3": "Related display logic but changes here would not fix the underlying iteration problem"
    }}
}}
\end{lstlisting}
\end{tcolorbox}

\begin{tcolorbox}[promptbox={Prompt 6: ROUND 1 MODIFICATION LOCATION PROMPT}]
\begin{lstlisting}[basicstyle=\ttfamily\footnotesize, breaklines=true]
You are an expert software engineer tasked with identifying specific code locations that need to be modified to solve a given issue.

**Issue Description:**
{issue_description}

**Selected Localization Chain:**
{chain_info}

**Your Task:**
Analyze the localization chain and identify the specific locations within this chain that need to be modified to solve the issue. Focus on pinpointing the exact functions, methods, or code blocks that require changes.

**CRITICAL REQUIREMENT FOR INSTRUCTIONS:**
- Each suggested_approach must be a DETAILED, STEP-BY-STEP instruction
- Include specific code examples, parameter names, and implementation details
- Specify exact lines to modify, functions to add, and variables to change
- Provide concrete implementation guidance that a developer can directly follow
- Include error handling, edge cases, and validation requirements
- Mention specific imports, dependencies, or setup needed

**Instructions:**
1. Examine each entity in the localization chain and its code
2. Identify which specific parts of the code are causing the issue or need enhancement
3. Determine the precise locations where modifications should be made
4. Explain why each location needs modification and what type of change is required
5. Prioritize the modifications by importance (most critical first)
6. For each modification, provide DETAILED implementation instructions with specific code examples

**Output Format:**
Return a JSON object with your analysis:
```json
{{
    "modification_locations": [
        {{
            "entity_id": "specific_entity_id",
            "location_description": "Specific function/method/lines that need modification",
            "modification_type": "fix_bug|add_feature|refactor|optimize",
            "priority": "high|medium|low",
            "reasoning": "Detailed explanation of why this location needs modification",
            "suggested_approach": "DETAILED step-by-step implementation instructions with specific code examples, parameter names, exact function signatures, error handling, and complete implementation guidance that can be directly executed by a developer"
        }}
    ],
    "overall_strategy": "Overall approach to solving the issue using these modifications",
    "confidence": 85
}}
```

**Example of DETAILED suggested_approach:**
Instead of: "Add proper termination condition"
Provide: "Modify the __iter__ method in the Paginator class by adding a counter variable 'current_page = 1' at the beginning. Then add a while loop condition 'while current_page <= self.num_pages:' to replace the infinite loop. Inside the loop, yield 'self.page(current_page)' and increment 'current_page += 1'. Add try-catch block to handle PageNotAnInteger and EmptyPage exceptions by catching them and breaking the loop. Import the exceptions 'from django.core.paginator import PageNotAnInteger, EmptyPage' at the top of the file."
\end{lstlisting}
\end{tcolorbox}

\begin{tcolorbox}[promptbox={Prompt 7: ROUND 2 COMPREHENSIVE MODIFICATION PROMPT}]
\begin{lstlisting}[basicstyle=\ttfamily\footnotesize, breaklines=true]
"""
You are an expert software engineer participating in a collaborative code review process to determine the best approach for solving a software issue.

**Issue Description:**
{issue_description}

**Selected Localization Chain:**
{chain_info}

**Your Initial Analysis:**
{your_initial_analysis}

**Other Agents' Analyses:**
{other_agents_analyses}

**Your Task:**
Based on the issue, the localization chain, your initial analysis, and insights from other agents, provide a refined and comprehensive analysis of where and how the code should be modified.

**CRITICAL REQUIREMENT FOR REFINED INSTRUCTIONS:**
- Each suggested_approach must be EXTREMELY DETAILED with complete implementation guidance
- Include specific code snippets, exact function signatures, and parameter details
- Provide line-by-line modification instructions where applicable
- Specify all necessary imports, dependencies, and setup requirements
- Include comprehensive error handling and edge case considerations
- Mention testing requirements and validation steps
- Provide specific examples of input/output or before/after code states

**Instructions:**
1. Review your initial analysis and the analyses from other agents
2. Identify common patterns and disagreements in the proposed modifications
3. Synthesize the best insights from all analyses
4. Refine your modification recommendations based on collective wisdom
5. Provide a more comprehensive and well-reasoned final recommendation
6. Ensure each suggested_approach contains exhaustive implementation details

**Output Format:**
Return a JSON object with your refined analysis:
```json
{{
    "refined_modification_locations": [
        {{
            "entity_id": "specific_entity_id",
            "location_description": "Specific function/method/lines that need modification",
            "modification_type": "fix_bug|add_feature|refactor|optimize",
            "priority": "high|medium|low",
            "reasoning": "Enhanced reasoning incorporating insights from other agents",
            "suggested_approach": "EXHAUSTIVE step-by-step implementation guide including: exact code snippets to add/modify/remove, complete function signatures, all required imports, parameter validation, error handling, edge cases, testing considerations, and specific examples of before/after states",
            "supporting_evidence": "References to other agents' insights that support this decision"
        }}
    ],
    "overall_strategy": "Comprehensive strategy refined through collaborative analysis",
    "confidence": 90,
    "key_insights_learned": "What you learned from other agents' analyses",
    "potential_risks": "Potential risks or challenges identified through collaborative review"
}}
```

Remember: Each suggested_approach should be so detailed that a developer can implement it without additional research or clarification.
\end{lstlisting}
\end{tcolorbox}

\begin{tcolorbox}[promptbox={Prompt 8: FINAL DISCRIMINATOR PROMPT}]
\begin{lstlisting}[basicstyle=\ttfamily\footnotesize, breaklines=true]
You are the lead software architect making the final decision on a code modification plan. Multiple expert engineers have provided their analyses for solving a software issue.

**Issue Description:**
{issue_description}

**Selected Localization Chain:**
{chain_info}

**All Agents' Final Analyses:**
{all_agents_analyses}

**Your Task:**
Synthesize all the expert analyses and create a definitive, actionable modification plan that will solve the issue effectively and safely.

**CRITICAL REQUIREMENTS FOR INSTRUCTIONS:**
- Every instruction MUST be a concrete modification action (Add, Remove, Modify, Replace, Insert, etc.)
- NO verification, checking, or validation instructions (avoid "Verify", "Ensure", "Check", "Maintain", etc.)
- Each instruction should specify exactly WHAT to change and HOW to change it
- Focus on direct code modifications that implement the solution

**Instructions:**
1. Analyze all the expert recommendations and identify the most reliable and consistent suggestions
2. Resolve any conflicts between different expert opinions using technical merit
3. Create a prioritized, step-by-step modification plan with ONLY concrete modification actions
4. Ensure the plan is practical, safe, and addresses the root cause of the issue
5. Include specific instructions for each modification
6. The output context should be as detailed as possible
7. Use action verbs like: "Add", "Modify", "Replace", "Insert", "Update", "Change", "Remove", "Implement"

**Output Format:**
Return a comprehensive modification plan:
```json
{{
    "final_plan": {{
        "summary": "High-level summary of the modification approach",
        "modifications": [
            {{
                "step": 1,
                "instruction": "Concrete modification instruction using action verbs (Add/Modify/Replace/etc.)",
                "context": "File path and specific location (e.g., function, method, line range)",
                "type": "fix_bug|add_feature|refactor|optimize",
                "priority": "critical|high|medium|low",
                "rationale": "Why this modification is necessary and how it contributes to solving the issue",
                "implementation_notes": "Specific technical details for implementation"
            }}
        ],
        "execution_order": "The recommended order for implementing these modifications",
        "testing_recommendations": "Suggested testing approach for validating the modifications",
        "risk_assessment": "Potential risks and mitigation strategies"
    }},
    "confidence": 95,
    "expert_consensus": "Summary of areas where experts agreed",
    "resolved_conflicts": "How conflicting expert opinions were resolved"
}}
```

**Examples of GOOD instructions:**
- "Add maxlength attribute to the widget configuration"
- "Modify the widget_attrs method to include max_length parameter"
- "Replace the current field initialization with max_length support"
- "Insert validation logic for maximum length"

**Examples of BAD instructions (DO NOT USE):**
- "Verify the max_length setting" 
- "Ensure proper validation"
- "Check if the field is configured correctly"
- "Maintain the existing functionality"

Focus on creating a plan that can be directly executed by a modification agent with clear, actionable steps.
\end{lstlisting}
\end{tcolorbox}

\end{document}